# Magnetic and transport properties of Sb$_2$Te$_3$ doped with high concentration of Cr

Sachin Gupta[1,2], Shun Kanai[3,4,5], Fumihiro Matsukura,[2,3,4,5*], and Hideo Ohno[1,2,3,4,5]

[1]*Center for Innovative Integrated Electronic Systems, Tohoku University, 468-1 Aramaki Aza Aoba, Aoba-ku, Sendai 980-0845, Japan*

[2]*WPI Advanced Institute for Materials Research, Tohoku University, 2-1-1 Katahira, Aoba-ku, Sendai 980-8577, Japan*

[3]*Laboratory for Nanoelectronics and Spintronics, Research Institute of Electrical Communication, Tohoku University, 2-1-1 Katahira, Aoba-ku, Sendai 980-8577, Japan*

[4]*Center for Spintronics Research Network, Tohoku University, 2-1-1 Katahira, Aoba-ku, Sendai 980-8577, Japan*

[5]*Center for Spintronics Integrated Systems, Tohoku University, 2-1-1 Katahira, Aoba-ku, Sendai 980-8577, Japan*

*Electronic mail: f-matsu@wpi-aimr.tohoku.ac.jp

We report on molecular beam epitaxy and properties of a magnetic topological insulator, Cr doped Sb$_2$Te$_3$. The composition analysis reveals that Cr replaces Sb site, and x-ray diffraction confirms that single phase textured crystal structure can be obtained for (Cr$_x$Sb$_{1-x}$)$_2$Te$_3$ with $x$ up to 0.44. Further increase in $x$ results in phase separation or precipitates in the material. The Curie temperature $T_C$ increases with $x$ up to 0.44, and reaches to 250 K, which is the highest $T_C$ observed till now in magnetically doped topological insulators.

Topological insulators (TIs) have become a focus of interest in condensed matter physics due to their intriguing properties resulting from gapped (insulating) bulk states and gapless (conducting) surface states. In the surface states, carrier spins are coupled with their momenta and are protected by the time reversal symmetry (TRS).[1] These unique surface states have been anticipated to exhibit various quantum phenomena which have potential in contributing to the developments in low power spintronics.[1,2] TRS in these materials can be broken by doping magnetic impurity or by ferromagnetic proximity effect, which results in the opening of band gap at the Dirac point.[3,4] Ferromagnetism in TIs results in many exotic physical phenomena such as quantum anomalous Hall effect and magneto-electric effects.[5,6] These properties stimulated interest in magnetically doped TIs for unlocking further novel quantum phenomena to pave the way for future spintronics.

Magnetic TIs have been synthesized by doping nonmagnetic TIs with 3$d$ transition metals such as V, Cr, Mn, Fe, and Cu. The doping induces various magnetic properties such as ferromagnetism, antiferromagnetism, and superconductivity.[7–13] To elucidate the nature of magnetism, one needs to vary the transition metal concentration as well as carrier concentration in the material. It has been observed that in the bulk matrix, the solubility of transition metal is rather low at the equilibrium growth condition.[14] Researchers have found that the low temperature molecular beam epitaxy (MBE) allows us to dope higher concentration of transition metal because of its non-equilibrium growth nature, which helps to enhance the Curie temperature $T_C$.[8] Among all the ferromagnetic TIs, Cr doped $Sb_2Te_3$, $(Cr_xSb_{1-x})_2Te_3$ has been reported to show ferromagnetism at a high $T_C$ of 190 K at Cr composition $x$ of ~0.3.[8] $Sb_2Te_3$ has a tetradymite crystal structure (space group $R\bar{3}m$) with quintuple (Te-Sb-Te-Sb-Te) layers piled up along $c$-axis via van der Waals interaction.[15]. In order to further probe the composition dependence of crystallographic, magnetic as well as electrical properties of $(Cr_xSb_{1-x})_2Te_3$, we study MBE growth of $(Cr_xSb_{1-x})_2Te_3$ and their properties by varying $x$ up to 0.61 and show that many properties of this system are quite similar to those of (Ga,Mn)As.

We grow $(Cr_xSb_{1-x})_2Te_3$ films with different $x$ on a semi-insulating GaAs (111)B substrate by MBE. The substrate is deoxidized by heating it up to 700°C inside the MBE chamber with base pressure < 8×10$^{-8}$ Pa, and then cooled down to growth temperature fixed at 280°C. The beam equivalent pressure (BEP) ratio between Sb and Te is kept to be ~1:7. We vary $x$ by tuning Cr/Sb beam flux ratio by changing their source temperatures. The *in situ* reflection high energy electron diffraction (RHEED) shows a streaky pattern throughout the growth, indicating two-dimensional growth of single crystals. After the growth, the film thickness $t$ (typically between 50 and 60 nm) is measured by cross-sectional scanning electron microscopy. The Cr composition $x$ is determined by energy dispersive x-ray spectroscopy (EDX). The $\theta$-$2\theta$ x-ray diffraction (XRD) measurements are performed using Cu K$\alpha$ radiation. The magnetic measurements as a function of temperature and magnetic field are carried out using a vibrating sample magnetometer. Transport measurements are performed using van der Pauw method.

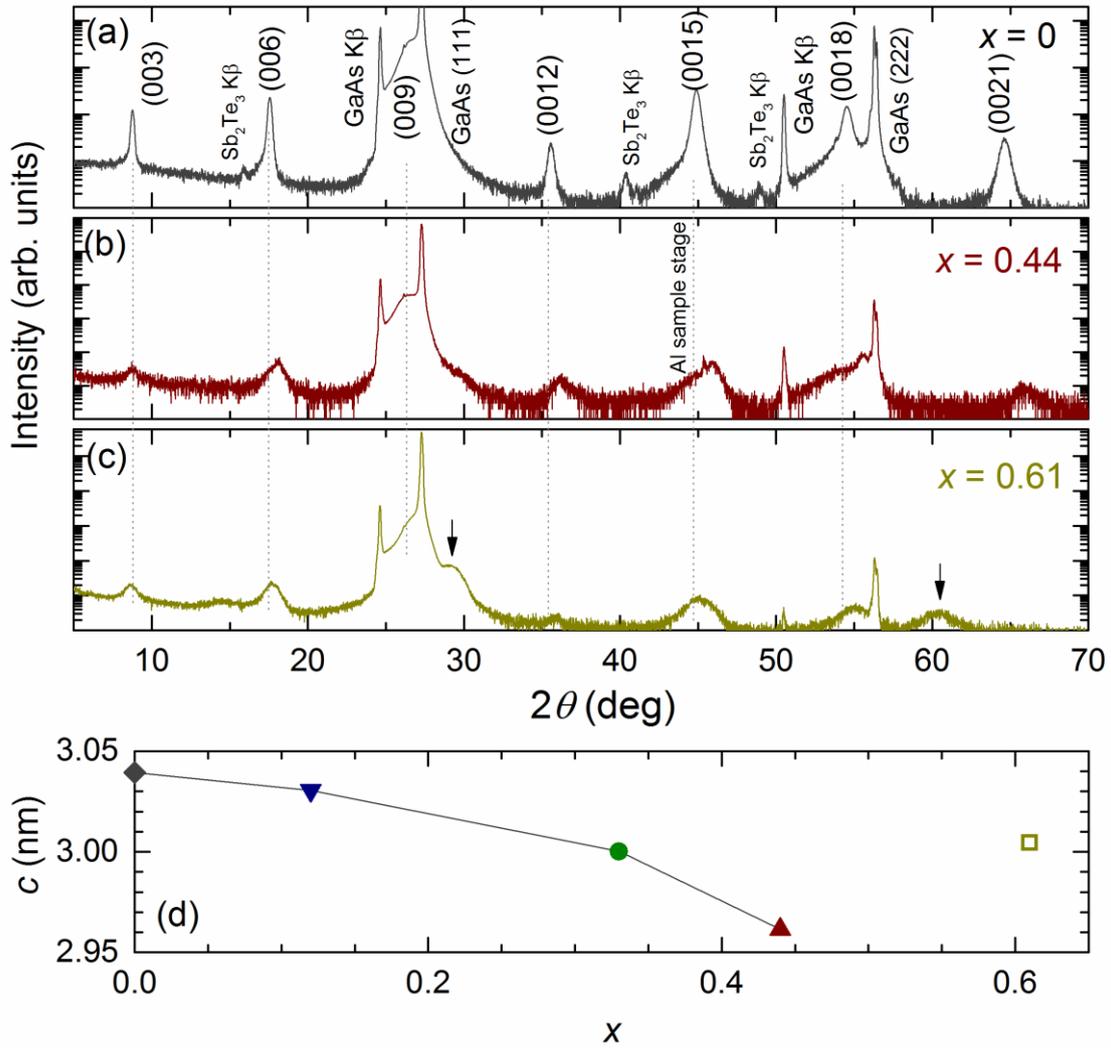

**Fig. 1** (a) $\theta$-$2\theta$ x-ray diffraction (XRD) patterns for $(Cr_xSb_{1-x})_2Te_3$ films with (a) Cr composition $x = 0$, (b) $x = 0.44$, and (c) $x = 0.61$. Vertical dotted lines are guide to see slight shift in peak positions on varying Cr concentration $x$. Arrows in (c) indicate additional peak positions. (d) The $c$-axis lattice parameter as a function of $x$, in which filled symbols for $(Cr,Sb)_2Te_3$ with uniform Cr distribution and unfilled symbol for with phase separation or precipitates.

The result of EDX indicates that we obtain (Cr,Sb)$_2$Te$_3$ with the nominal highest $x$ of 0.61. The ratio of ([Cr]+[Sb]):[Te] is nearly 2:3, indicating that Cr atoms replace Sb. The XRD pattern of the (Cr,Sb)$_2$Te$_3$ films up to $x = 0.44$ shows (003) family peaks along with GaAs (111) substrate peaks, which confirms tetradymite structure with growth direction along $c$-axis (Fig. 1(a)). The $c$-axis lattice parameter for Sb$_2$Te$_3$ determined to be 3.04 nm from Bragg's law using the peak position, which is close to the reported value.[15] The peaks slightly shift to higher $2\theta$ angle on increasing $x$ up to 0.44 as shown in Fig. 1(b), indicating the decrease of $c$-axis lattice parameter, which is consistent with the previous report.[8] The sample with $x = 0.61$ shows almost the same peak positions with those for the (Cr,Sb)$_2$Te$_3$ with $x = 0.33$ and additional peaks at $2\theta \sim 30°$ and $60°$ (marked by arrows) as shown in Fig. 1(c). This suggests phase separation or precipitation in (Cr,Sb)$_2$Te$_3$ with high $x$. The Cr composition dependence of $c$-axis lattice parameter is summarized in Fig. 1(d).

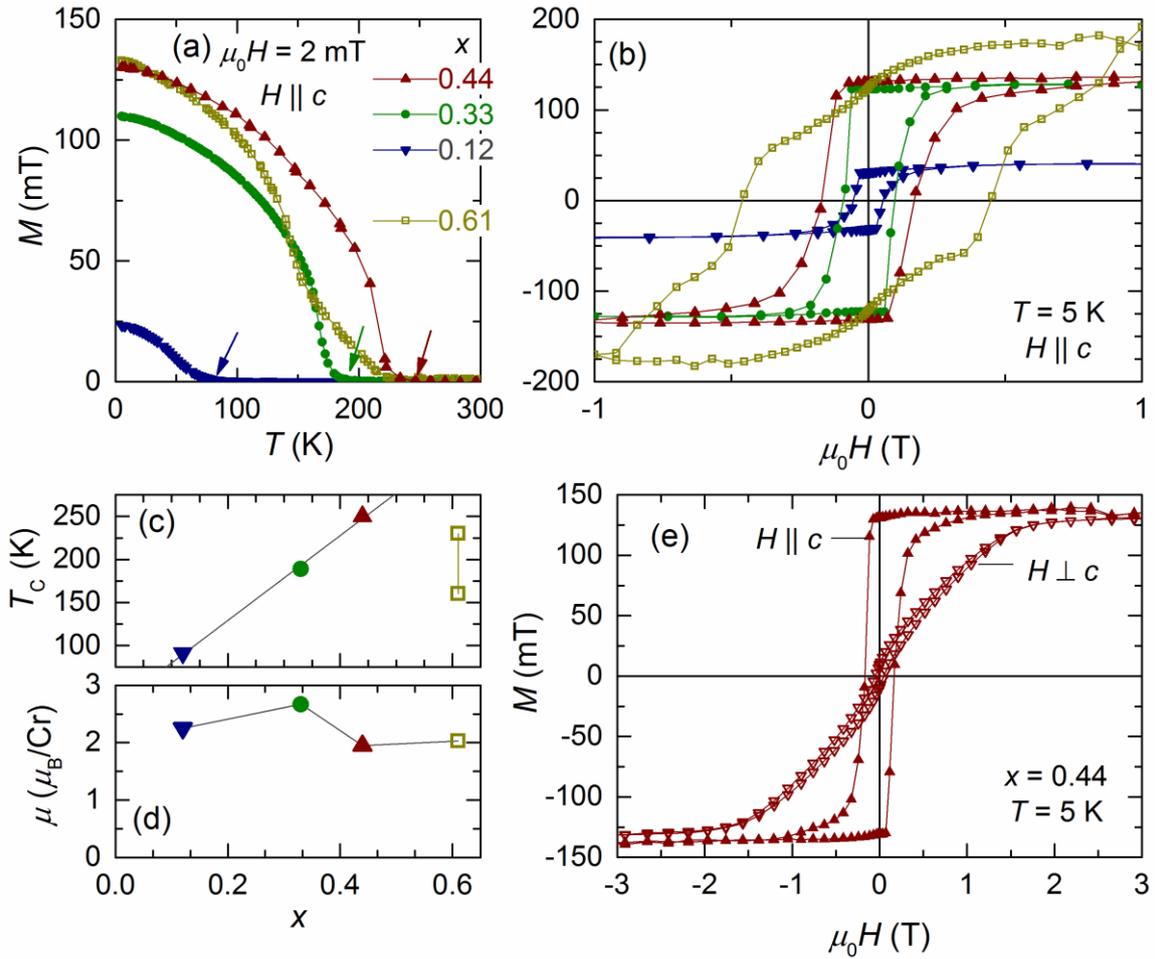

**Fig. 2** (a) The temperature $T$ dependence of magnetization $M$ of (Cr$_x$Sb$_{1-x}$)$_2$Te$_3$ films with different Cr compositions *x* after cooled in perpendicular magnetic field of 0.1 T. Arrows mark the Curie temperature $T_C$. (b) Magnetization curves under perpendicular magnetic fields at 5 K. (c) The Curie temperature $T_C$ as a function of *x*. Solid line is linear fit for $(Cr,Sb)_2Te_3$ with $x < 0.5$. (d) Magnetic moments $\mu$ per Cr as a function of *x*. (e) Magnetization curves for $(Cr,Sb)_2Te_3$ with $x = 0.44$ at 5 K under perpendicular and in-plane magnetic fields.

Figure 2(a) shows the temperature dependence of magnetization *M*. The sample is cooled down from room temperature to the lowest measured temperature (5 K) in the presence of 0.1 T applied normal to the film plane, and then data are recorded while heating the sample to 400 K under 2 mT, a field to compensate the remanent field in the magnetometer.[16] Upon heating, magnetization decreases monotonically and shows magnetic phase transition from ferromagnetic to paramagnetic at $T_C$ ($T_C$ is marked with an arrow). On increasing *x*, the magnetization increases and $T_C$ shifts to higher temperatures reaching to 250 K for $(Cr,Sb)_2Te_3$ with $x = 0.44$. Further increase of *x* results in a curve that has two or more components (phases) as indicated by a rapid drop of *M* at ~160 K followed by a gradual decrease to ~230 K, which appears to be consistent with the XRD results.

Figure 2(b) shows the magnetization curves for $(Cr,Sb)_2Te_3$ with different *x* at 5 K under perpendicular magnetic fields, in which the diamagnetic contribution from the substrate is subtracted from the total magnetization using a linear fit in high fields. A clear hysteresis loop with coercive field $H_C$ of ~53, ~91, and ~167 mT, and saturation magnetization $M_S$ of ~39, ~130 and ~135 mT for $x = 0.12$, 0.33, and 0.44, respectively. The values of $M_S$ and $H_C$ increase with increasing *x*. The sample with $x = 0.61$ shows hysteresis with steps, indicating the presence of two ferromagnetic phases as detected by the temperature dependence of *M*. To see how $T_C$ varies with *x*, we plot $T_C$ vs. *x* in Fig. 2(c), in which $T_C$ increases almost linearly with *x* up to 0.44, similar to what was reported in (Ga,Mn)As.[17,18] The observation also appears to be consistent with theoretical calculation, which shows the ferromagnetic interaction is mediated by a Te anion between Cr atoms as well as carriers.[19] The magnetic moment $\mu$ per Cr atom determined from $M_S$ is between $2\mu_B$ and $3\mu_B$ (Fig. 2(d)), where $\mu_B$ is the Bohr magneton, showing that Cr is electrically neutral ($Cr^{3+}$) or acts as a donor ($Cr^{4+}$). Figure 2(e) shows the magnetization curves for $(Cr,Sb)_2Te_3$ with $x = 0.44$ at 5 K under fields applied along out-of-plane and in-plane directions. The film has an easy axis of magnetization parallel to the *c*-axis and hard axis of magnetization in *ab*-plane like previous reports on ferromagnetic TIs.[8] The anisotropy field is as large as ~2 T.

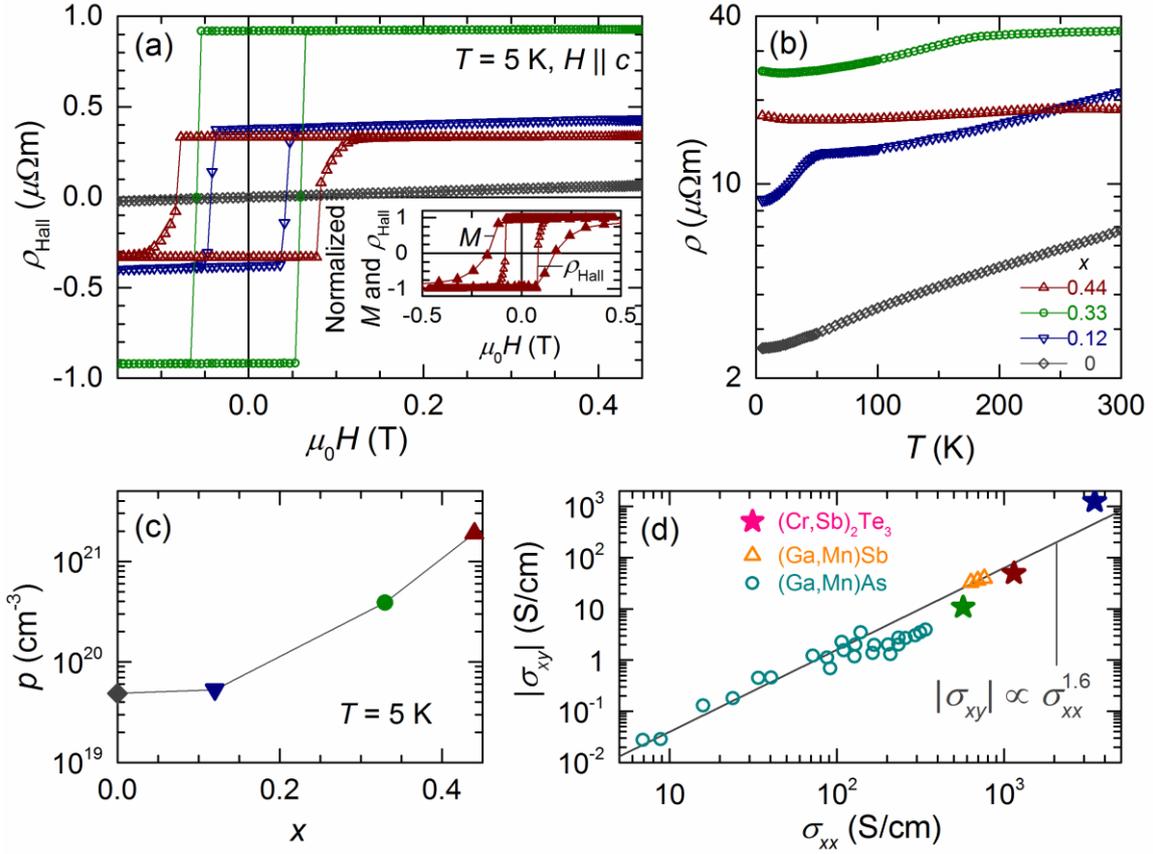

**Fig. 3** (a) Hall resistivity $\rho_{Hall}$ at 5 K and (b) temperature $T$ dependence of longitudinal resistivity $\rho$ of $(Cr,Sb)_2Te_3$ as a function of Cr composition $x$. (c) Cr composition $x$ dependence of hole concentration $p$. (d) Double-logarithm plots of Hall conductivities $|\sigma_{xy}|$ versus longitudinal conductivities $\sigma_{xx}$ for $(Cr_xSb_{1-x})_2Te_3$ (filled stars) along with those for (Ga,Mn)As and (Ga,Mn)Sb (unfilled symbols).[23] Solid line represents the empirical scaling relation $|\sigma_{xy}| \propto \sigma_{xx}^{1.6}$.

Figures 3(a) and 3(b) display the field dependence of the Hall resistivity $\rho_{Hall}$ at 5 K and the temperature dependence of the longitudinal resistivity $\rho$ in zero magnetic field for $(Cr,Sb)_2Te_3$ with different $x$, respectively. The $\rho_{Hall}$ in magnetic materials can be expressed by the following expression; $\rho_{Hall} = R_0\mu_0H + R_SM$ with the ordinary Hall coefficient $R_0$, the anomalous Hall coefficient $R_S$, and permeability in the free space $\mu_0$. $(Cr,Sb)_2Te_3$ with $x = 0$ shows the ordinary Hall effect (OHE) as shown by the linear dependence of $R_{Hall}$ with $H$, while $(Cr,Sb)_2Te_3$ with $x = 0.12$-$0.44$ show the anomalous Hall effect (AHE) due to magnetic contribution. The positive slope of $\rho_{Hall}$ in high field region, where AHE saturates, indicates that hole type carriers are dominated in these films possibly due to native defects such as Sb antisites acting as acceptors.[20] The hole concentration $p$ in $(Cr,Sb)_2Te_3$ determined from the slope is summarized in Fig. 3(c), which suggests that the introduction of Cr promotes the formation of electrically active defects and results in the increase of hole concentration. The inset of Fig. 3(a) shows the magnetic-field dependence of $\rho_{Hall}$ and $M$ measured at 5 K for the film $x = 0.44$. Hysteresis loop in $\rho_{Hall}$ is more square than $M$. Same behavior was reported in (Ga,Mn)As.[21] The difference in the two hysteresis loops arises from the difference in the two measurements probe the sample. In magnetic measurements, $M$ comes from spins from whole regime, while in transport measurement $\rho_{Hall}$ detects magnetization from the conducting region, where current flows easily than less conducting region.[21] However, we obtain the same $T_C$ from the Arrott plots using Hall resistance data as that determined from the magnetization data.

Figure 3(d) shows double-logarithm plots of Hall conductivities $\sigma_{xy}$ versus longitudinal conductivities $\sigma_{xx}$ (stars) along with those obtained previously for (Ga,Mn)As and (Ga,Mn)Sb.[22,23] The plots obey an empirical scaling relation, $|\sigma_{xy}| \propto \sigma_{xx}^{1.6}$, as observed for a number of materials including typical ferromagnetic semiconductors.[24] As seen in Fig. 3(b), below 25 K, $\rho$ for the Cr doped films starts increasing upon further cooling. This low temperature rise is more pronounced as $x$ increases. This feature can be explained by two mechanisms: One is the freezing of bulk carriers at low temperatures, where bulk becomes insulating and surface states dominate.[25] The other is weak localization.[26] We observe negative magnetoresistance at low temperatures, which suggests that the low temperature increase in $\rho$ is due to weak localization in TIs.[27] Again, similar behavior was observed for (Ga,Mn)As.[17,26] Both magnetic and transport properties of $(Cr,Sb)_2Te_3$ are thus quite similar to those reported for (Ga,Mn)As, which suggests that $(Cr,Sb)_2Te_3$ is another useful material to investigate the spin-orbit coupling related phenomena in magnetic materials.[28] In addition to this, the bulk conduction in the system under study can be suppressed by applying electrical gating or chemical substitution. This can provide information of the mechanism of the observed ferromagnetism,[29] and make this high $T_C$ material a potential candidate for applications in low power spintronics.

In summary, we grow $(Cr_xSb_{1-x})_2Te_3$ with different $x$ on a semi-insulating GaAs (111)B substrate using molecular beam epitaxy. Structural, magnetic, and transport properties do not indicate any secondary phase for $(Cr,Sb)_2Te_3$ with $x$ up to 0.44. Magnetic and transport properties are very similar to those of (Ga,Mn)As, and show robust ferromagnetic ordering in these films. The Curie temperature increase with increase in Cr composition $x$, and reaches to ~250 K for $(Cr,Sb)_2Te_3$ with $x = 0.44$. Further increase in $x$ results in the phase separation or precipitates in the material, and does not lead further enhancement of the Curie temperature.

## Acknowledgements

The authors thank T. Dietl, B. Jinnai, A. Okada, and S. Miyakozawa for useful discussion and technical help. The work was supported in part by a Grant-in-Aid for Scientific Research from MEXT (#26103002), Program on Open Innovation Platform with Enterprises, Research Institute and Academia (OPERA), JST, the JSPS Core-to-Core program, and the Cooperative Research Project Program of RIEC.